\documentclass[preprint,showpacs,preprintnumbers,amsmath,amssymb]{revtex4-1}

\usepackage{graphicx}
\usepackage{dcolumn}
\usepackage{bm}
\usepackage{caption}
\usepackage{subcaption}

\usepackage{dcolumn}
\usepackage{bm}

\newcommand{\be}{\begin{equation}}
\newcommand{\ee}{\end{equation}}
\newcommand{\bse}{\begin{subequations}}
\newcommand{\ese}{\end{subequations}}
\newcommand{\bary}{\begin{eqnarray}}
\newcommand{\eary}{\end{eqnarray}}
\newcommand{\bwt}{\begin{widetext}}
\newcommand{\ewt}{\end{widetext}}

\begin{document}


\title{Hadronic-Origin orphan TeV flare from the  1ES 1959+650}
\author{Sarira Sahu$^{*}$,  Andres Felipe Osorio Oliveros$^{**}$, Juan
  Carlos Sanabria$^{**}$  
}
\affiliation{
$^{*}$Instituto de Ciencias Nucleares, Universidad Nacional Aut\'onoma de M\'exico, 
Circuito Exterior, C.U., A. Postal 70-543, 04510 Mexico DF, Mexico\\
$^{**}$Universidad de Los  Andes, Bogota, Colombia
}


\begin{abstract}

The 1ES 1959+650 is a high-peaked BL Lacertae object. On  4th of June, 2002, it
exhibited a  strong TeV  flare without any low energy
counterpart, providing for the first time an example of an {\it orphan}
flare from a blazar. Observation of
this orphan flare is in striking disagreement with the predictions
of the leptonic models thus challenging the conventional 
synchrotron self-Compton (SSC)  interpretation of the TeV emission.
Here we propose that, the
low energy tail of the  SSC photons in the blazar jet serve as the
target for the Fermi-accelerated high energy protons of energy $\lesssim 100$
TeV, within the jet to produce the  
TeV photons through the decay of neutral pions from the delta
resonance. Our model explains very nicely the observed TeV flux
from this orphan flare
and we also estimate the high energy neutrino flux from this flaring event.

\end{abstract}

\maketitle

\section{Introduction}

Blazars are a subclass of  active galactic nuclei (AGNs) which include
flat-spectrum radio quasars (FSRQs) and BL Lacertae (BL Lac)
objects. Both FSRQs and BL Lac are characterized by strong and rapid
flux variability across the entire electromagnetic spectrum which are
predominantly nonthermal. The emission extends all the way from radio to
$\gamma$-ray and is believed to be produced in a highly relativistic
plasma jet pointing along the line of sight to the observer. 
Due to the small viewing angle of the jet, it is possible to observe the
strong relativistic effects, such as the boosting of the emitted power
and a shortening of the characteristic time scales, as short as
minutes\cite{Abdo:2009wu,Aharonian:2007ig}. Thus these objects are important to study the energy
extraction mechanisms from the central super-massive black holes, 
physical properties of the astrophysical jets,
the acceleration mechanisms of the charged particles in the jet and production of
ultra high energy cosmic rays.
The spectral energy distribution (SED) of blazars is characterized by
two nonthermal bumps \cite{von Montigny:1995yz,Fossati:1998zn} and depending on the location of the first peak
of the SED, these are often sub-classified into low energy peaked
blazars (LBLs) and high energy peaked blazars (HBLs)\cite{Padovani:1994sh}. In LBLs, the
first peak is in the near-infrared/optical energy range and for HBLs
it is in the UV or X-rays range, while the second peak is around GeV
energy range for LBLs and for HBLs it is in the TeV energy range.

There is a general consensus that the low energy peak is produced due
to the synchrotron emission from accelerated electrons and positrons
in the emitting region. Although, the origin of the high energy peak
remains inconclusive, the leading interpretation is the 
SSC model, where  the high
energy emission is from a population of relativistic electrons
up scattering their self-produced synchrotron photons\cite{Ghisellini:1998it} or by external
photons. This model is found very successful
in explaining the multi-wavelength emission from BL Lac objects and FR I
galaxies such as NGC 1275 and M87
\cite{Ghisellini:1998it,Fossati:1998zn} and also Centaurus A\cite{Abdo:2010fk,Roustazadeh:2011zz}.
The inevitable outcome of the leptonic models is that,
flaring at TeV energy should be accompanied by a simultaneous flaring
in the synchrotron peak. Alternatively,
in the hadronic models, the high energy peak is produced due to 
proton synchrotron emission or decay of neutral pions formed in
cascades from the interaction of high energy proton beam
with the radiation or gas clouds surrounding the
source \cite{Atoyan:2001xy}. In this scenario, a strong correlation between the gamma-ray
and the neutrino fluxes is expected \cite{Reimer:2005sj,Halzen:2005pz,AdrianMartinez:2011jz}.

The AGN 1ES 1959+650  was first detected in the Einstein IPC Slew Survey\cite{Elvis:1992}
and classified as a HBL subclass, based on its
X-ray to radio flux ratio\cite{Schachter:1993}. 
It has a redshift of $z=0.047$\cite{VeronCetty:2006zz}  with a luminosity distance of
$d_L=210$ Mpc and the mass of the central black hole is
estimated to be $\sim 1.5\times 10^8 M_{\odot}$. 
Because of the HBL subclass and low redshift, it had long been
considered a potential candidate for TeV $\gamma$-ray source.
The first very high
energy (VHE) gamma-ray from 1ES 1959+650 was observed in 1998 by the
Seven Telescope Array in Utah, with a $3.9\,\sigma$ significance\cite{Nishiyama:1999js}
and later on other observations were also reported but the
observed flux was weak in both gamma-rays and in X-rays. The HEGRA
collaboration reported only a marginal signal during its observations
from 2000 until early 2002. In May 2002, 1ES 1959+650  underwent a
strong TeV outburst which was observed by Whipple\cite{Holder:2002ru} and HEGRA
experiments\cite{Aharonian:2003be} as well as in the X-ray range by RXTE experiments. The
X-ray flux smoothly declined  throughout the following month. However,
during this smooth decline period, a second TeV flare was observed after few
days (on 4th of June) of the initial one without a X-ray counterpart\cite{Krawczynski:2003fq}. 
On the other hand, In July 2006, the BL Lac PKS 2155-304 had a giant
TeV flare recorded by HESS\cite{Aharonian:2007ig} which was
accompanied by an increase in X-ray flux and can be explained through
SSC model.  So the observation of
the ``orphan'' flare in 1ES 1959+650  is in striking disagreement with the predictions
of the leptonic models thus challenging the SSC interpretation of the TeV emission.
Non observation of a significant X-ray activity could naturally be
interpreted by the suppression of electron acceleration and inverse
Compton scattering as production mechanism for very high energy (VHE) gamma rays in favor
of hadronic models.  Motivated by the above argument, a hadronic synchrotron mirror model was
proposed by B\"ottcher\cite{Bottcher:2004qs} to explain this orphan
TeV flare and also  the neutrino flux is estimated during the flaring\cite{Reimer:2005sj}. In
this model, the flare is explained through the decay of neutral pions
to gamma rays when the former are produced due to the interaction of
high energy cosmic ray (HECR) protons with the primary synchrotron photons that have
been reflected off clouds located at a few pc above the accretion
disk. These photons are blue shifted in the jet frame so that there
will be substantial decrease in the HECR proton energy to overcome the
threshold for delta resonance and,  at the same time, it is an
alternative to the standard scenario where HECR protons interact with
the synchrotron photons, where one needs HECR protons to be Fermi
accelerated to high energy.  But the idea of the reflection of
synchrotron photons off-cloud at a few pc above the accretion
disk demands a special geometry of the jet+cloud system, which plays a vital
role. Also how efficiently these photons will be reflected from the
cloud is rather unclear.

\section{The Hadronic model}

In a recent paper Sahu and Zhang\cite{Sahu:2012wv} have
shown that the multi-TeV emission from the AGN, Centaurus A, detected by
HESS during 2004 to 2008 can be well interpreted as the decay of
neutral pions from the $\Delta$-resonance of $p\gamma$ interactions of Fermi-accelerated high
energy protons in the jet with the seed photons around the second SED
peak at $\sim$170 keV.   It is also shown that this same model is
consistent with the detection of two ultra-high energy cosmic ray
events by Pierre Auger Observatory from the Centaurus A
direction. After the success of the hadronic model to explain the
multi-TeV photon flux from Centaurus A\cite{Sahu:2012wv} , we would like to apply it to
the orphan TeV flare of 1ES1959+650. Here we assume the standard
interpretation of the leptonic model to explain both, low and high 
energy peaks, by synchrotron and SSC photons respectively as in the
case of Centaurus A.  
Thereafter, we propose that the low energy tail of the SSC photons in the jet
serve as the target for the Fermi-accelerated high energy protons to produce the pions
through delta resonance and their subsequent decay to high energy
photons. In this scenario the condition of  B\"ottcher\cite{Bottcher:2004qs} is
automatically satisfied without the need of a special geometry. 

The SED of the 1ES1959+650 is fitted quite well with the leptonic one-zone
synchrotron and SSC model\cite{Tagliaferri:2008qk,Gutierrez:2006ak}. In this model, the emitting region is a blob
with comoving radius $R'_b$  moving with a velocity
$\beta_c$ corresponding to a bulk Lorentz
factor $\Gamma$ and seen at
an angle $\theta_{ob}$ by an observer which results with a Doppler
factor ${\cal D}=\Gamma^{-1} (1-\beta_c \cos\theta_{ob})^{-1}$.  The emitting region is filled with an
isotropic electron population and a randomly oriented magnetic field
$B'$. The electrons have a power-law spectrum given as $dN/dE\propto
E^{-\alpha}$ with the power index $\alpha \ge 2$.

The energy spectrum of the Fermi-accelerated protons in the blazar jet
is also assumed to be of power-law. Due to high radiative losses, electron
acceleration is limited. On the other hand, protons and heavy nuclei can
reach UHE through the same acceleration mechanism. 

The pion production in $p\gamma$ collision through $\Delta$-resonance is
\be
p+\gamma \rightarrow \Delta^+\rightarrow  
 \left\{ 
\begin{array}{l l}
 p\,\pi^0, & \quad \text {fraction~ 2/3}\\
  n\,\pi^+  \rightarrow n e^{+}\nu_e\nu_\mu \bar\nu_\mu, 
& \quad  \text {fraction~ 1/3}\\
\end{array} \right. ,
\label{decaymode}
\ee
which has a cross section $\sigma_{\Delta}\sim 5\times 10^{-28}\,
{\rm cm}^2$. The charged $\pi$'s subsequently decay to charged leptons and
neutrinos, while neutral $\pi$'s decay to GeV-TeV photons.
For the above process to take place, the center-of-mass energy of the
interaction has to exceed the $\Delta$-mass 1.232 GeV which  corresponds
to the kinematical condition
\be
E'_p = \frac{(m^2_{\Delta}-m^2_p)} {2 \epsilon'_\gamma (1-\beta_p
  \cos\theta)}
\simeq \frac{0.32\,{\rm  GeV}^2}{\epsilon'_\gamma },
\ee
where $E'_p$ and $\epsilon'_\gamma$ are
the proton and the background  photon energies in the comoving frame
of the jet, respectively (quantities with a prime are in the comoving
frame and without prime are in the observer frame). Also for high energy protons we take
$\beta_p\simeq 1$. Since in the comoving frame the protons collide
with the SSC photons
from all directions, in our calculation we consider an average value $(1-\cos\theta) \sim 1$ 
($\theta$ in the range of 0 and $\pi$). 
In the observer frame, one can
re-write the matching condition as
\be
E_p \epsilon_\gamma \simeq 0.32~ \frac{\Gamma {\cal D}}{(1+z)^2} ~{\rm GeV}^2~.
\label{resonant1} 
\ee
Here 
\be
\epsilon_\gamma = \frac{{\cal D} \epsilon'_\gamma}{(1+z)},
\ee
is the observed background photon energy, while
\be
E_p=\frac{\Gamma E'_p}{(1+z)}, 
\ee
is the energy of the proton as measured by the observer on Earth,
if it could escape the source and reach earth without energy loss.

In the comoving frame, each pion carries $\sim 0.2$ of the proton energy.
Considering that each $\pi^0$ decays into two $\gamma$-rays, the $\pi^0$-decay
$\gamma$-ray energy in the observer frame ($E_{\gamma}$)can be written as
\be
E_\gamma = \frac{1}{10}\frac{{\cal D}}{(1+z)} E'_p = \frac{\cal
  D}{10\,\Gamma} E_p. 
\label{EgammaEp}
\ee
The matching condition between the $\pi^0$-decay photon energy
$E_{\gamma}$ and the 
target photon energy $\epsilon_{\gamma}$ is therefore
\be
E_\gamma \epsilon_\gamma \simeq 0.032~\frac{ {\cal D}^2}{(1+z)^2} ~{\rm GeV}^2.
\label{Eegamma} 
\ee

With the leptonic one-zone synchrotron and SSC interpretation, different models
use different parameters\cite{Tagliaferri:2008qk,Gutierrez:2006ak} to fit the SED of 1ES 1959+650. In all these
models, although the blob size differ by about 1 to 2 orders of
magnitudes ($1.4\times 10^{14}\,{\rm cm} \le R'_b \le 1.4\times
10^{16}\,{\rm cm})$,
the bulk Lorentz factor $\Gamma$ and ${\cal D}$ are almost the same
($18 \le \Gamma\simeq {\cal D}\le 20$) and the comoving 
magnetic field $B'$ lies in the range 0.04 G to 0.25 G.
The multiwavelength observation of the SED of  1ES
1959+650 was performed in 2006 May and fitted with the above one-zone
model by Tagliaferri et. al \cite{Tagliaferri:2008qk} (their FIG: 8), where the
parameters used for the jet are
$\Gamma\simeq {\cal D}=18$, $R'_b =7.3\times 10^{15}$ cm and $B'=0.25$
G. 

The time averaged TeV energy spectrum (above 1.4 TeV) of the flaring state of the 1ES
1959+650 was well fitted with pure power-law by the HEGRA
collaboration and the power-law
spectral index is $\alpha=2.83\pm 0.14_{\rm stat} \pm 0.08_{\rm
  sys}$ or by a power-law with an exponential cut-off at
$E_c=(4.2^{+0.8}_{-0.6{\rm stat}}\pm 0.9_{\rm sys})$ TeV and a
spectral index of $1.83\pm0.15_{\rm stat}\pm 0.08_{\rm sys}$\cite{Aharonian:2003be,Daniel:2005rv}

\begin{figure}[t!]
\vspace{0.3cm}
{\centering
\resizebox*{0.3\textwidth}{0.25\textheight}
{\includegraphics{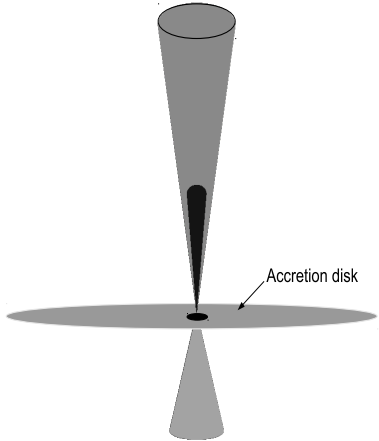}}
\par}
\caption{Geometry of the orphan flaring of blazar 1ES 1959+650: the interior
  compact cone (jet) is responsible for the orphan flaring and the exterior cone
  corresponds to the normal jet. 
}
\label{blaz}
\end{figure}

In this work we assume  that the flaring occurs within a  compact and confined
volume of radius $R'_f$ inside the blob of radius $R'_b$ ($R'_f <
R'_b$). The geometrical description of the jet structure in the orphan
flare is shown in
FIG. 1. This double jet structure may be applicable to all orphan flares. 
In this scenario the internal and the external jets are moving with the
same bulk Lorentz factor 
$\Gamma$ and the Doppler factor ${\cal  D}$ as the blob. 
Within the confined volume, the injected spectrum of the
Fermi accelerated charged particles have 
a power-law with an exponential cut-off, and for the
protons with energy $E_p$ it is given as
as 
\be
\frac{dN_p}{dE_p}\propto  E_p^{-\alpha} e^{-E_p/E_{p,c}},
\label{powerlaw}
\ee
where the high energy proton has the cut-off energy $E_{p,c}$ and the
spectral index $\alpha > 2$. Also in this small volume, the comoving
photon number density 
$n'_{\gamma, f}$ (flaring) is much higher than rest of the blob
$n'_{\gamma}$ (non-flaring),  which can be due to the copious annihilation of electron positron
pairs, splitting of photons in the magnetic field, enhance IC
photons in this region and Poynting flux dominated flow
which can form, from the magnetic  reconnection in the strongly
magnetized plasma around the base of the jet\cite{Giannios:2009kh,Giannios:2009pi}. This can be expressed as
${n'_{\gamma, f}(\epsilon_\gamma)}=\lambda {n'_{\gamma}(\epsilon_\gamma)}$,
where $\lambda \gg 1$. So the ratio of photon densities at two different
background energies $\epsilon_{\gamma_1} $  and $\epsilon_{\gamma_2} $
in flaring and non-flaring states remains the same, that is
\be
\frac{n'_{\gamma, f}(\epsilon_{\gamma_1})}
{n'_{\gamma, f}(\epsilon_{\gamma_2})}=\frac{n'_\gamma(\epsilon_{\gamma_1})}
{n'_\gamma(\epsilon_{\gamma_2})}.
\ee 
The high energy protons will collide with the  low-energy tail of the
SSC photons within the confined volume of radius $R'_f$ in the energy range $\sim 1$ MeV to 8 MeV ($\sim 2.0\times
10^{20}$ Hz to $\sim 2.0\times 10^{21} $ Hz) to produce
$\Delta$-resonance. It can be observed from Eqs. (\ref{EgammaEp}) and
(\ref{Eegamma}) that, in the hadronic model, high energy $\gamma$-rays are
produced when very high energy protons  collide with
low-energy SSC photons and vice versa and the 
optical depth of the $\Delta$-resonance
process is given as
\be
\tau_{p\gamma}=n'_{\gamma, f}\sigma_{\Delta} {\cal R}'_f.
\label{optdep}
\ee 
The comoving photon number density within the confined volume can be
given in terms of the luminosity $L_{\gamma}$ as 
\be
n'_{\gamma, f} = \eta \frac{L_\gamma}{{\cal D}^{2+\kappa}} \frac{(1+z)}{4\pi
{\cal R'}^2_f \,\epsilon_\gamma},
\ee
with $\kappa \sim (0-1)$
(depending on whether the jet is continuous or discrete) and $\eta
\sim 1$ .
In the flaring region, the number of $\pi^0$-decay
photons at a given energy depend on the number of high energy protons and
the optical depth, i.e. 
$N(E_\gamma)\propto N(E_p)/\tau_{p\gamma} \propto N(E_p) n'_{\gamma, f}(\epsilon_\gamma)$,
where $E_\gamma$, $E_p$ and $\epsilon_\gamma$ satisfy the
$\Delta$-resonance matching conditions given in Eqs.(\ref{resonant1})
and (\ref{Eegamma}) and  the $\gamma$-ray flux from $\pi^0$ decay will be 
\bary
F_{\gamma}(E_{\gamma}) &\equiv& E^2_{\gamma} \frac{dN(E_\gamma)}{dE_\gamma} \nonumber\\
&\propto & E^2_p \frac{dN(E_p)}{dE_p} n'_{\gamma, f}(\epsilon_\gamma).
\eary
So the high energy observed photon flux from $\pi^0$-decay at two different
observed photon energies $E_{\gamma,1}$ and $E_{\gamma,2}$ will scale as
\bary
\frac{F_\gamma(E_{\gamma_1})}{F_\gamma(E_{\gamma_2})} &=& \frac{n'_{\gamma, f}(\epsilon_{\gamma_1})}
{n'_{\gamma, f}(\epsilon_{\gamma_2})}
\left(\frac{E_{\gamma_1}}{E_{\gamma_2}}\right)^{-\alpha+2}
e^{-(E_{\gamma_1}-E_{\gamma_2})/E_c}\nonumber\\
&=& \frac{n'_\gamma(\epsilon_{\gamma_1})}
{n'_\gamma(\epsilon_{\gamma_2})}
\left(\frac{E_{\gamma_1}}{E_{\gamma_2}}\right)^{-\alpha+2}
e^{-(E_{\gamma_1}-E_{\gamma_2})/E_c}~,
\label{spectrum}
\eary
where $E_{\gamma_{1,2}}$ correspond to the proton energy $E_{p_{1,2}}$ and we
have used the relations $E_{p_1}/E_{p_2}=E_{\gamma_1}/E_{\gamma_2}$, and
$E_{p_{1,2}}/E_{p,c}=E_{\gamma_{1,2}}/E_c$. By using the known flux at a particular energy
in the flaring state, we can calculate the  flux at other energies using Eq.(\ref{spectrum}).

Out of $\tau^{-1}_{p\gamma}$ many protons, one interact with the SSC
background to produce photons and neutrinos as shown in
Eq.(\ref{decaymode}). So the fluxes of the TeV photons and the
Fermi accelerated high energy protons $F_p$, are related through
\be
F_p(E_p)=5\times\frac{3}{2} \frac{1}{\tau_{p\gamma}(E_p)}
F_{\gamma}(E_{\gamma}),
\label{protonflux}
\ee
where the factor 5 corresponds to 20\% of the proton energy 
taken by each $\pi^0$ and 3/2 is due to the 2/3 probability of $\Delta$-resonance
decaying to $p\pi^0$. 
Like photons, the proton fluxes at different energies $E_{p,1}$ and
$E_{p,2}$, scale as
\be
\frac{F_p(E_{p_1})}{F_p(E_{p_2})}=\left ( \frac{E_{p_1}}{E_{p_2}}
\right)^{-\alpha+2} e^{-(E_{p_1}-E_{p_2})/E_{p,c}}.
\ee

The fluxes of $\pi^+$ and $\pi^0$ are
related, because each pion carries 20\% of the proton energy, while
each neutrino and each $e^+$ carries 1/4 of the $\pi^+$ energy,  from the $\pi^0$
decay the photon carries 1/2 of the $\pi^0$ energy. The neutrino and
$e^+$, each has
energy $E_{\nu}=E_{e^+}=E_{\gamma}/2$ and the neutrino flux can be calculated
from the GeV-TeV photon flux, through
\be
F_{\nu}=\frac{3}{8} F_{\gamma},
\ee
where we assume that the TeV photon flux in the flaring state is solely due to
the hadronic process.

\begin{figure}[t!]
\vspace{0.3cm}
{\centering
\resizebox*{0.5\textwidth}{0.3\textheight}
{\includegraphics{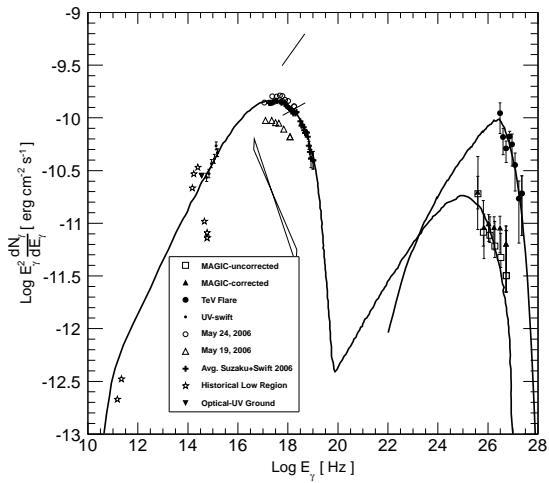}}
\par}
\caption{The observed SED  
$E^2_{\gamma}\frac{dN_{\gamma}}{dE_{\gamma}}$ (or $F_\gamma$) from
radio to $\gamma$-ray of 
 the blazar 1ES 1959+650  as measured at the end of 2006 May with
 other historical data from Ref. \cite{Tagliaferri:2008qk}. Different symbols are observations with 
  different sources marked within the box, and the curves are model fits:
  The continuous curve  (from low energy to high energy) is synchrotron + SSC fit from Tagliaferri
   et. al. \cite{Tagliaferri:2008qk}, while 
    the second continuous curve (extreme right) is the hadronic model fit to TeV flux from  
  $\pi^0$-decay. The X-ray spectra are the single-power-law fit taken
  from Ref. \cite{Krawczynski:2003fq}.
}
\label{SED}
\end{figure}

\section{Results}

During the flaring event, the high energy $\gamma-$rays flux was
observed in the energy range $1.26\, {\rm TeV} (3.05\times 10^{26} {\rm Hz})\lesssim E_{\gamma} \lesssim 9.4\, {\rm
  TeV} (2.3\times 10^{27} {\rm Hz}) $ by HEGRA and Whipple experiments. In the context of the
hadronic model that we consider here, the corresponding proton energy
will lie in the range $12\, {\rm TeV} \lesssim E_{p} \lesssim 94\, {\rm TeV}$
which will collide with the background photons in the energy range
$7.5\, {\rm MeV} (1.8\times 10^{21}\, {\rm Hz})\gtrsim
\epsilon_{\gamma} \gtrsim 1\, {\rm MeV} (2.4\times 10^{20}\, {\rm
  Hz})$ and these $\epsilon_{\gamma}$ lie exactly in the low energy tail of
the SSC photons as shown in FIG: 2,
calculated using the one-zone leptonic model.  Due to our ignorance of 
$n'_{\gamma,f}$, we can use the $n'_{\gamma}$
from the one-zone SSC fit to non-flaring state and use it in
Eq.(\ref{spectrum}) to calculate the flaring flux.

For $1\, {\rm MeV} \lesssim
\epsilon_{\gamma} \lesssim 7.5\, {\rm MeV}$, the corresponding photon density
in the blob is $300\, {\rm cm^{-3}} \gtrsim n'_{\gamma}(=n'_{\gamma,f}/\lambda) \gtrsim 88\,{\rm
  cm^{-3}}$.  Protons with energy $E_p < 12$ TeV will collide with
photons having energy $\epsilon_{\gamma} > 7.5$ MeV  and $n'_{\gamma,f}/\lambda
< 88\, {\rm cm^{-3}}$, to produce $E_{\gamma} < 1.2 $ TeV. Similarly, high energy
protons ($E_p > 94$ TeV) will collide with the low energy photons 
($\epsilon_{\gamma} < 1$ MeV with $n'_{\gamma,f}/\lambda > 300\,{\rm cm^{-3}}$) to produce
$E_{\gamma} > 9.4$ TeV. Here we use $\kappa=0$ and  for discrete jet, $n'_{\gamma,f}$
will be reduced by a factor of $ {\cal D}^{-1}$.

For the calculation of the TeV flux, first we take into account one of
the observed flaring fluxes with its corresponding energy for
normalization e.g. 
$F_{\gamma}(E_{\gamma_2}=1.26\, TeV)\simeq 10^{-10}\,{\rm erg}\,{\rm  cm}^{-2}\,{\rm
  s}^{-1})$
and  $n'_{\gamma}(\epsilon_{\gamma_2}=7.5
MeV)/\lambda\simeq 88\, {\rm cm^{-3}}$ and using it calculate the flux for other energies
with the Eq.(\ref{spectrum}). This we have done for
different observed fluxes for a better fit.
The spectral index
$\alpha$ and the cut-off energy $E_c$  are the free parameters in the
model and  the best fit is obtained for $\alpha=2.83$ and $E_c=5.0$
TeV\cite{Aharonian:2003be}. The $\gamma$-ray cut-off
energy of 5 TeV corresponds to $E_{p,c}=50$ TeV and above the cut-off
energy the flux decreases rapidly.

The Eddington luminosity of the blazar with the central black hole mass $\sim
1.5\times 10^8\,M_{\odot}$ is $L_{\rm Edd}\sim 2\times 10^{46} {\rm
  erg\, s^{-1}} (M/10^{8.3} M_{\odot})$ having the corresponding flux
$F_{\gamma}\sim 3.8\times 10^{-9}\, {\rm erg\, cm^{-2}\,s^{-1}}$. 
In the flaring state, the proton luminosity  $L_p(E_p=94\,{\rm TeV})$
has to be smaller than $L_{\rm Edd}$, which gives
$\tau_{p\gamma} \gtrsim 0.02$.
By considering $R'_f \sim 10^{14}$ cm, it gives
$n'_{\gamma,f} \gtrsim 4\times 10^{11}{\rm cm^{-3}}$. So the photon
density  has to be very high within the compact region.

The high energy protons will be accompanied by electrons in
the same energy range ($12\, {\rm TeV} \lesssim E_{e} \lesssim 94\,
{\rm TeV}$ ). These electrons will radiate synchrotron photons in the
jet magnetic field,  in the range $1.9\, {\rm MeV} (4.5\times 10^{20}\,
{\rm Hz}) \lesssim E_{\gamma} \lesssim 115.4\, {\rm MeV} (2.8\times
10^{22}\, {\rm Hz})$ which lies just in the lower part of
the SSC spectrum where the photon flux is quite low, as shown in FIG: 2. So even if the
flux in this region increases by an order of magnitude due to
electron synchrotron emission, there will still be a dip in the
spectrum which is unobserved.

During the flaring the high energy $\gamma$-rays were in the 
range 1.26 TeV $\lesssim E_{\gamma} \lesssim$ 9.4 TeV. If these photons were to interact with the
photons in the jet background to have $e^+e^-$ pair creation, then
each electron/positron will have an energy  $E_{\gamma}/2$. 
The $e^+$ produced during the $\pi^+$ decay as shown in
Eq.(\ref{decaymode}) will also carry same amount of energy as the
individual $e^-$ and $e^+$ in the pair creation process.
Within the
jet, these electrons and positrons may undergo synchrotron emission
where the magnetic field is about 0.25 G as discussed earlier. We
found that the synchrotron photons will be in the range $1.3\times
10^{18}$ Hz (0.005 MeV) to $7\times 10^{19}$ Hz (0.29 MeV), which lies
in the high-energy edge of the synchrotron spectrum as shown in FIG. 2. However the mean
free path $\lambda_{\gamma\gamma}=(n'_{\gamma}
\sigma_{\gamma\gamma})^{-1}$  for the pair creation in the jet background is 
several orders of magnitude larger than blob radius
$R'_b$.  Even if we replace $n'_{\gamma}$ by the photon density in the
flaring region i.e. $n'_{\gamma,f}$, the photon mean free path is
still much larger than the blob radius. This implies that TeV photons
will not be degraded by pair creation and therefore no synchrotron
emission will take place for the above $e^+e^-$ pairs. But the $e^+$
created due to pion decay will emit synchrotron radiation in the
frequency band $1.3\times
10^{18}$ Hz  to $7\times 10^{19}$ Hz, and its flux
will be much smaller than $F_{\gamma}(E_{\gamma}=1.26\, {\rm TeV})/8$ i.e. $F_{e^+,{\rm syn}}\ll
1.25\times 10^{-11}\, {\rm erg}\,{\rm cm}^{-2}\,{\rm s}^{-1}$. This is
much below the observed synchrotron flux in the normal
case as can be seen from FIG. 2 and can't  be observed  during
the flaring event.  The non-observation of $e^+$ synchrotron flux
during the flaring is also explicitly shown  by
B\"ottcher which makes the flare genuinely orphan\cite{Bottcher:2004qs}.
It is true that the synchrotron radiation of the positrons
from the $\pi^+$ decay 
and the electrons which accompany the Fermi accelerated protons will
make the dip shallower which will be 
$\ll 1.25\times 10^{-11}\, {\rm erg}\,{\rm cm}^{-2}\,{\rm s}^{-1}$  in the  range $\sim 10^{18}$ Hz to
$\sim 10^{22}$ Hz. This is the frequency range wich falls in the falling
edge of the synchrotron spectrum and the tail of the IC spectrum. 
It is important to note that the marginal enhancement in the photon flux in
the above frequency range is only during the flaring event and not in the 
normal circumstances when leptonic process is the sole contributor to
the SED.

The TeV flux from 1ES 1959+650 can in principle be reduced due to
the absorption of TeV photons by the diffuse extragalactic background
radiation through 
$\gamma_{\rm TeV}+\gamma_{\rm b}\rightarrow e^+e^-$ process. But the
energy range of our interest $\sim 1$ TeV to several TeV, the spectral
shape remains nearly unchanged due to the almost constant optical
depth for most of the extragalactic background
radiation\cite{Aharonian:2003be}.

Our result is shown in FIG. 2, which fits very well with the
observed flaring flux. Also it is observed that the flux increases
for  $E_{\gamma} < 1.2$ TeV due to the  high proton flux in this
energy range. However, in order not to violate the Eddington
luminosity, the proton energy spectrum must break to a harder index
(e. g. $\alpha \sim 2.3$) below 12 TeV. Here we have introduced a break
at $E_{p,b}\sim 12$ TeV below which $\alpha=2.3$ and above this energy
$\alpha=2.83$. The spectral energy distribution $F_{\gamma}$ falls
below $E_{p,b}\sim 12$ TeV as shown in FIG. 2.
It may also so happen that 
$\epsilon_{\gamma} > 7.5$ MeV are very much suppressed
within the compact region implying
$n'_{\gamma}$ to be  too low for
$\Delta$-resonance to occur  hence $E_{\gamma} < 1.2$ TeV  production
is negligible. The flux decreases rapidly for
$E_{\gamma} > 9.4$ TeV because of the exponential cut-off.
This corresponds to a proton flux above 94 TeV which will also fall
exponentially as shown in Eq.(\ref{powerlaw}).

We have also estimated  the neutrino flux from decay of the charged
pions during the intense flare, where the neutrino  flux will be 
$4.5\times 10^{-9}\, {\rm GeV}\,{{\rm cm}^{-2}}\,{\rm s}^{-1} \lesssim F_{\nu}
\lesssim 2.6\times 10^{-8}\,{\rm GeV}\,{\rm {cm}^{-2}\,s^{-1}}$ corresponding to neutrino
energy in the range $4.7\, TeV \gtrsim  E_{\nu}\gtrsim
0.6\, TeV$. \\

\section{Conclusions}

We have employed the hadronic model to interpret the TeV  emission
from the orphan flaring event of June 2002, from the blazar 1ES 1959+650. In this picture,
Fermi-accelerated protons of energy $\lesssim\,100$ TeV interact with the low energy $\sim(1-8)$ MeV tail of the SSC
photons  in a very compact and confined region
of the jet to produce $\Delta$-resonance and its subsequent decay to
photons through neutral pion decay. The TeV photons thus obtained
are proportional to both, high energy proton spectrum which is a
power-law with an exponential cut-off and the low energy
SSC photon density in the blazar jet.  Our result fits very well with
the observed flux from the flaring event. We have also estimated the
neutrino flux from this event. 

We are thankful to Bing Zhang for useful discussions.  
S.S. is thankful to Departamento de Fisica de Universidad de los
Andes, Bogota, Colombia, for their kind hospitality during his several visits.
This work is partially supported by DGAPA-UNAM (Mexico) Project 
No. IN103812 and IN105211.


\begin{thebibliography}{55}
\expandafter\ifx\csname natexlab\endcsname\relax\def\natexlab#1{#1}\fi
\expandafter\ifx\csname bibnamefont\endcsname\relax
  \def\bibnamefont#1{#1}\fi
\expandafter\ifx\csname bibfnamefont\endcsname\relax
  \def\bibfnamefont#1{#1}\fi
\expandafter\ifx\csname citenamefont\endcsname\relax
  \def\citenamefont#1{#1}\fi
\expandafter\ifx\csname url\endcsname\relax
  \def\url#1{\texttt{#1}}\fi
\expandafter\ifx\csname urlprefix\endcsname\relax\def\urlprefix{URL }\fi
\providecommand{\bibinfo}[2]{#2}
\providecommand{\eprint}[2][]{\url{#2}}

\bibitem{Abdo:2009wu} 
  A.~A.~Abdo {\it et al.}  [Fermi LAT Collaboration],
  Astrophys.\ J.\  {\bf 700}, 597 (2009)
  [arXiv:0902.1559 [astro-ph.HE]].

\bibitem{Aharonian:2007ig} 
  F.~Aharonian,
  Astrophys.\ J.\  {\bf 664}, L71 (2007)
  [arXiv:0706.0797 [astro-ph]].

\bibitem{von Montigny:1995yz} 
  C.~von Montigny, D.~L.~Bertsch, J.~Chiang, B.~L.~Dingus, J.~A.~Espositio, C.~E.~Fichtel, J.~M.~Fierro and R.~C.~Hartman {\it et al.},
  Astrophys.\ J.\  {\bf 440}, 525 (1995).

\bibitem{Fossati:1998zn}
  G.~Fossati, L.~Maraschi, A.~Celotti, A.~Comastri and G.~Ghisellini,
  Mon.\ Not.\ Roy.\ Astron.\ Soc.\  {\bf 299} (1998) 433
  [arXiv:astro-ph/9804103].

\bibitem{Padovani:1994sh} 
  P.~Padovani and P.~Giommi,
  Astrophys.\ J.\  {\bf 444}, 567 (1995)
  [astro-ph/9412073].

\bibitem{Ghisellini:1998it}
  G.~Ghisellini, A.~Celotti, G.~Fossati, L.~Maraschi and A.~Comastri,
  Mon.\ Not.\ Roy.\ Astron.\ Soc.\  {\bf 301} (1998) 451
  [arXiv:astro-ph/9807317].

\bibitem{Abdo:2010fk}
  A.~A.~Abdo {\it et al.}  [Fermi LAT Collaboration],
  Astrophys.\ J.\  {\bf 719}, 1433-1444 (2010).
  [arXiv:1006.5463 [astro-ph.HE]].

\bibitem{Roustazadeh:2011zz}
  P.~Roustazadeh and M.~B\"ottcher,
  Astrophys.\ J.\  {\bf 728}, 134 (2011).


\bibitem{Atoyan:2001xy} 
  A.~M.~Atoyan, P.~M.~Chadwick, M.~K.~Daniel, K.~Lyons, T.~J.~L.~McComb, J.~M.~McKenny, S.~J.~Nolan and K.~J.~Orford {\it et al.},
Astron.\ Astrophys.\  {\bf 383}, 864 (2002)
  [astro-ph/0112177].

\bibitem{Reimer:2005sj} 
  A.~Reimer, M.~Bottcher and S.~Postnikov,
  Astrophys.\ J.\  {\bf 630}, 186 (2005)
  [astro-ph/0505233].


\bibitem{Halzen:2005pz} 
  F.~Halzen and D.~Hooper,
  Astropart.\ Phys.\  {\bf 23}, 537 (2005)
  [astro-ph/0502449].

\bibitem{AdrianMartinez:2011jz} 
  S.~Adrian-Martinez, I.~A.~Samarai, A.~Albert, M.~Andre, M.~Anghinolfi, G.~Anton, S.~Anvar and M.~Ardid {\it et al.},
  arXiv:1111.3473 [astro-ph.HE].

\bibitem{Elvis:1992} 
 M.~Elvis {\it et al.},
 Astrophys..\ J.\ Supp.\ S.\ {\bf 80}, 257 (1992).

\bibitem{Schachter:1993}
J.~F.~Schachter {\it et al.},
Astrophys.\ J.\  {\bf 412}, 541 (1993).

\bibitem{VeronCetty:2006zz} 
  M.~-P.~Veron-Cetty and P.~Veron,
  Astron.\ Astrophys.\  {\bf 455}, 773 (2006).

\bibitem{Nishiyama:1999js} 
  T.~Nishiyama {\it et al.}  [Utah Seven Telescope Array Collaboration],
  In Salt Lake City 1999, Cosmic ray, vol. 3,  370-373

\bibitem{Holder:2002ru} 
  J.~Holder {\it et al.}  [VERITAS Collaboration],
  Astrophys.\ J.\  {\bf 583}, L9 (2003)
  [astro-ph/0212170].

\bibitem{Aharonian:2003be} 
  F.~Aharonian {\it et al.}  [HEGRA Collaboration],
  Astron.\ Astrophys.\  {\bf 406}, L9 (2003)
  [astro-ph/0305275].

\bibitem{Krawczynski:2003fq} 
  H.~Krawczynski, S.~B.~Hughes, D.~Horan, F.~Aharonian, M.~F.~Aller, H.~Aller, P.~Boltwood and J.~Buckley {\it et al.},
  Astrophys.\ J.\  {\bf 601}, 151 (2004)
  [astro-ph/0310158].


\bibitem{Bottcher:2004qs} 
  M.~B\"ottcher,
  Astrophys.\ J.\  {\bf 621}, 176 (2005)
  [Erratum-ibid.\  {\bf 641}, 1233 (2006)]
  [astro-ph/0411248].


\bibitem{Sahu:2012wv} 
  S.~Sahu, B.~Zhang and N.~Fraija,
  Phys.\ Rev.\ D {\bf 85}, 043012 (2012)
  [arXiv:1201.4191 [astro-ph.HE]].

\bibitem{Tagliaferri:2008qk} 
  G.~Tagliaferri and L.~Foschini,
  Astrophys.\ J.\  {\bf 679}, 1029 (2008)
  [arXiv:0801.4029 [astro-ph]].


\bibitem{Gutierrez:2006ak} 
  K.~Gutierrez {\it et al.}  [VERITAS Collaboration],
  Astrophys.\ J.\  {\bf 644}, 742 (2006)
  [astro-ph/0603013].

\bibitem{Daniel:2005rv} 
  M.~K.~Daniel {\it et al.}  [The VERITAS Collaboration],
  Astrophys.\ J.\  {\bf 621}, 181 (2005)
  [astro-ph/0503085].

\bibitem{Giannios:2009kh} 
  D.~Giannios, D.~A.~Uzdensky and M.~C.~Begelman,
Mon.\ Not.\ Roy.\ Astron.\ Soc.\  {\bf 395} (2009) L29
  [arXiv:astro-ph/9807317].
  arXiv:0901.1877 [astro-ph.HE].

\bibitem{Giannios:2009pi} 
  D.~Giannios, D.~A.~Uzdensky and M.~C.~Begelman,
Mon.\ Not.\ Roy.\ Astron.\ Soc.\  {\bf 402} (2010) 1649
  arXiv:0907.5005 [astro-ph.HE].


\end{thebibliography}
\end{document}